\documentclass[reprint, amsmath,amssymb, aps, pra, longbibliography]{revtex4-1}

\usepackage{soul}
\usepackage{amsmath}   
\usepackage{amssymb}
\usepackage{graphicx}
\usepackage{dcolumn}
\usepackage{bm}
\usepackage{amsfonts}
\usepackage{subfigure}
\usepackage{array}
\usepackage{float}
\usepackage{color}
\usepackage[colorlinks=true,linkcolor=blue]{hyperref}%
\hypersetup{allcolors=blue}
\usepackage[normalem]{ulem}
\usepackage{xcolor}

\newcommand{\ket}[1]{\ensuremath{\left\vert #1 \right\rangle}}

\begin{document}

\title{Dynamic polarizability of the $^{85}$Rb $5D_{3/2}$-state in 1064~nm light}
\date{\today }

\author{A. Duspayev}
    \email{alisherd@umich.edu}
\affiliation{Department of Physics, University of Michigan, Ann Arbor, MI 48109, USA}
\author{R. Cardman}
\affiliation{Department of Physics, University of Michigan, Ann Arbor, MI 48109, USA}
\author{G. Raithel}
\affiliation{Department of Physics, University of Michigan, Ann Arbor, MI 48109, USA}

\begin{abstract}
We report a measurement of the dynamic (ac) scalar polarizability of the $5D_{3/2}$ state in $^{85}$Rb atoms at a laser wavelength of 1064~nm. Contrary to a recent measurement in Phys. Rev. \textbf{A} 104, 063304 (2021), the experiments are performed in a low-intensity regime in which the ac shift is less than the $5D_{3/2}$ state's hyperfine structure, 
as utilized in numerous experiments with cold, trapped atoms. The extracted ac polarizability is $\alpha_{5D_{3/2}} = -499\pm59$~a.u., within the uncertainty of the aforementioned previous result. The calibration of the 1064~nm light intensity, performed by analyzing light shifts of the D1 line, is the main source of uncertainty. Our results are useful for applications of the Rb $5D_{3/2}$ state in metrology, quantum sensing, and fundamental-physics research on Rydberg atoms and molecules.          
\end{abstract}

\maketitle

\section{Introduction}
\label{sec:intro}
Precision measurements of atomic structure and properties are of major importance for a wide range of applications including optical atomic clocks~\cite{atomicclocksreview, bloom2014, martin2019}, quantum computing and simulations~\cite{Saffman.2016, morgado2021}, and field sensing~\cite{Sedlacek2012, Holloway2014, anderson2021, RFMS}. Atom-field interactions are an essential tool in a variety of these topics, as quite often it is required to trap neutral atoms, perform optical or microwave excitations and enhance or mitigate certain effects. Shifts of atomic energy levels due to the ac Stark effect~\cite{Delone, gerginov2018twoprap, martin2019frequency}, $\Delta W$, can be listed among such effects and are described (in terms of the local field intensity $I$) by

\begin{equation}
\Delta W = -\frac{\alpha_{\xi}(\lambda) I}{2 c \epsilon_0} ,
\label{eq:ACShift}
\end{equation}

\noindent where $\alpha_{\xi}(\lambda)$ is the ac polarizability of an atomic state $\ket{\xi}$ at the wavelength of the applied field, $\lambda$, $c$ is the speed of light, and $\epsilon_0$ is the vacuum permittivity. Determination of $\alpha_{\xi}(\lambda)$ is crucial for applications of certain atomic or molecular states that involve atom-field interactions~\cite{MarinescuPolar, Safronova2006, Topcu2013, LeKien2013} including laser cooling and trapping~\cite{metcalf}, quantum control~\cite{patsch, cardman2020} and optical clocks~\cite{atomicclocksreview, bloom2014, martin2019}.

In this paper, we report a measurement of $\alpha$ for the $5D_{3/2}$ state of $^{85}$Rb atom at $\lambda = 1064$~nm. This wavelength is widely used for optical dipole traps and lattices for both neutral atoms~\cite{grimm2000} and ions~\cite{weckesser} because it is a commonly available wavelength emitted at high optical powers from diode and Nd:YAG fiber lasers. A recent measurement at this wavelength yielded a result of $\alpha_{5D_{3/2}} = -524\pm17$ (in atomic units)~\cite{cardman2021}. The experiment in~\cite{cardman2021} was performed at an intensity on the order of 100~GW/m$^2$, which is high enough to decouple the hyperfine (HF) structure and to cause large broadening due to photoionization (PI), as 1064~nm is above the PI threshold of the $5D_{3/2}$ state (1251.52~nm). However, it is not always desirable~\cite{Saffman.2016} or even practical to conduct experiments under such high-intensity conditions. In the present work, we therefore perform two-photon spectroscopy of the ac Stark effect of the $5D_{3/2}$ state and measure $\alpha_{5D_{3/2}}$ at low intensity of the 1064-nm light, where HF interactions are on the same order as the light shifts. 

The $5D$ states of Rb are of general interest for several reasons. Two-photon transitions between the ground, $5S_{1/2}$, and $5D_J$ states are relatively strong, narrow and can be driven by lasers in readily accessible visible or NIR ranges. These features make the aforementioned transitions attractive in metrology and as frequency references~\cite{nez1993optcomoptical, touahri1997optcomfrequency, hilico1998epjapmetrological, bernard2000, terra2016apbultra, martin2019, quinn2003metpractical}. Moreover, Rb atoms in $5D_J$ states can be excited into Rydberg $nP$ and $nF$ states for studies on three-photon EIT~\cite{carr2012, Moore2019a}, Rydberg molecules~\cite{duspayev2021} and spectroscopy of high-angular-momentum Rydberg states~\cite{Younge2010, cardman2021njp}. For the Rb $5D_{3/2}$ state, the static polarizability~\cite{snigirev2014prameasurement}, the HF structure~\cite{nez1993optcomoptical, terra2019}, the radiative lifetime~\cite{sheng2008}, and, recently, the dynamic polarizability at 1064~nm~\cite{cardman2021} have been measured. In addition, the large PI cross-section of the $5D_{3/2}$ state~\cite{Aymar1984, duncan2001prameasurement, cardman2021} could be used for generation of ultracold plasmas~\cite{pohl2004, viray2020} and experiments on atom-ion interactions~\cite{schmid, seckerpra, ewald, dieterleprl}. 

Our paper is organized as follows: theoretical considerations of atom-light interactions and details of the experimental setup and the data analysis are provided in Sec.~\ref{sec:methods}. Results are presented and discussed in Sec.~\ref{sec:res}. The paper is concluded in Sec.~\ref{sec:concl}. 

\section{Methods}
\label{sec:methods}
\subsection{Theoretical background}
\label{subsec:theory}
Assuming a linearly polarized field, $\alpha_{\xi}$ in Eq.~\ref{eq:ACShift} can be expressed as

\begin{equation}
\alpha_{\xi}(\lambda) = \alpha^{(0)}_{\xi}(\lambda) + \eta_{\xi} \alpha^{(2)}_{\xi}(\lambda),
\label{eq:poltotal}
\end{equation}

\noindent where $\alpha^{(0)}_{\xi}$ and $\alpha^{(2)}_{\xi}$ are referred to as scalar and tensor dynamic polarizabilities, respectively, with the latter vanishing for states with $J < 1$. The elements of the second-rank tensor $\eta_{\xi}$ depend on the eigenstates of the Hamiltonian that describes the atom-field interaction. Circular polarization components would require an additional term proportional to the vector polarizability, $\alpha^{(1)}_{\xi}$, which is not included in Eq.~\ref{eq:poltotal} because in our experiment the light interacting with the atoms is linearly polarized.

\begin{figure}[t!]
 \centering
  \includegraphics[width=0.48\textwidth]{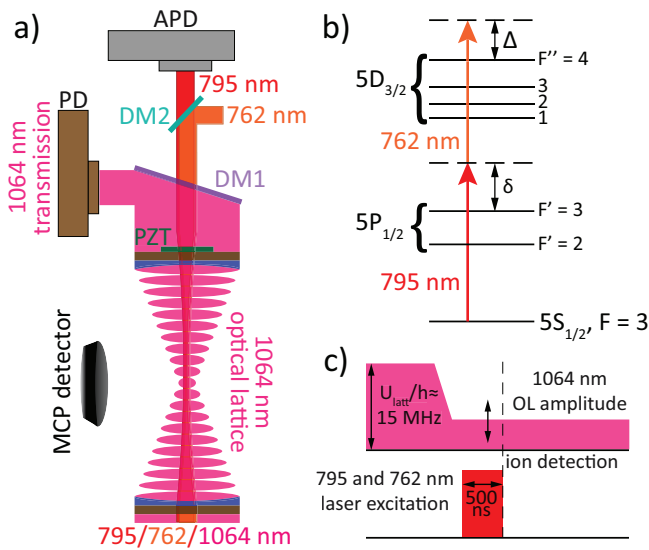}
  \caption{(Color online) Outline of experimental setup (a), utilized laser excitation scheme (not to scale) (b) and timing sequence of laser excitation and photoionization (c). ``APD" avalanche photo-detector; ``PD" photo-diode; ``DM" dichroic mirror; ``PZT" piezo-electric transducer. See text for details.} 
  \label{fig1}
\end{figure}

Depending on the strength of the applied field, the system can be dominated by the HF structure of the atom (weak-field case), the atom-light coupling (strong-field case) or reside in the intermediate regime. Comparing the system with an atom in an external magnetic field, the first two cases are analogous to the Zeeman and Paschen-Back regimes of the HF structure, respectively. The weak- and strong-field sets of the eigenstates are generally different. In the weak-field regime, the eigenstates of the Hamiltonian are given by the  $\{ \vert F, m_{F} \rangle \}$-basis, with HF quantum numbers $F$ and $m_F$.
An ac Stark interaction significantly larger than the HF interaction mixes the $\{ \vert F,m_{F} \rangle \}$ states, and the set of time-independent states approaches the $\{ \vert m_{J},m_{I} \rangle \}$-basis, with magnetic quantum numbers $m_J$ and $m_I$ denoting  electronic and nuclear spin along the lattice laser's electric-field direction. In the intermediate lattice-intensity regime the time-independent states transition from the low-field into the high-field basis~\cite{Chen2015,Neuzner2015}. Analytical expressions for the elements of $\eta_{\xi}$ can be derived in the weak- and strong-field cases (see, e.g.,~\cite{JamiePhD}). 
Under the presence of level broadening due to PI in the lattice-laser field, and if the tensor contribution of the polarizability is small, the transition of the time-independent states from the $\{ \vert F,m_{F} \rangle \}$ into the $\{ \vert m_{J},m_{I} \rangle \}$-basis may be blurred, as is the case for Rb~${5D_{3/2}}$ state in 1064-nm light.  A previous measurement of $\alpha_{5D_{3/2}}$~\cite{cardman2021} has been performed in the strong-field regime. In the present work, we conduct an experiment in the low-field regime. In the case that there are no unaccounted-for systematic 
errors in either of the measurements, the polarizabilities derived from the data should agree between low- and high-field measurements. The experimental challenges of the present low-field measurement and the methods employed to address them are described in Sec.~\ref{subsec:extr}.

\subsection{Experimental setup}
\label{subsec:expsetup}

Our measurements are performed using $^{85}$Rb atoms cooled within an intracavity optical lattice (OL). Detailed design and characterization of the utilized setup are provided in~\cite{Chen2014praatomtrapping}. The essential aspects of the apparatus are shown in Fig.~\ref{fig1}~(a). Rb atoms are cooled and trapped in a 3D magneto-optical trap (MOT), from which they are loaded into a vertically oriented OL formed using a TEM$_{00}$-mode of a near-concentric in-vacuum optical cavity. The OL laser has a wavelength of $\lambda =$ 1064~nm, for which $\alpha_{5D_{3/2}}$ is obtained. In addition to generating the $5D_{3/2}$ light-shifts to be measured, the lattice light also photo-ionizes the $5D_{3/2}$ atoms, which broadens the atomic levels and serves as a method for $\ket{5D_{3/2}}$-population readout via ion counting.

Two probe-laser beams at 795~nm and 762~nm (powers $\sim$1~$\mu$W and $\sim$50~$\mu$W before the cavity, respectively), are tuned by the means of two optical phase-locked loops (OPLLs) relative to fixed-frequency reference lasers that are locked to the relevant atomic transitions via saturated spectroscopy in Rb vapor cells.
The probe lasers are coupled to the cavity to drive atoms from the $\ket{5S_{1/2}, F=3}$ state to different HF states of the $5D_{3/2}$ level via the intermediate $\ket{5P_{1/2}}$ state, as shown in the level diagram in Fig.~\ref{fig1}~(b) (magnetic quantum numbers suppressed in the kets). Throughout the data acquisition, the 795~nm laser is held at a fixed frequency with a detuning of $\delta = 0.9$~GHz with respect to field-free $\ket{5S_{1/2}, F = 3} \rightarrow \ket{5P_{1/2}, F' = 3}$ transition. We scan the 762~nm laser frequency in 250-kHz steps to excite the Rb atoms into different HF states, $\ket{5D_{3/2}, F''}$. The detuning of the two-photon transition from the $\ket{5S_{1/2}, F = 3} \rightarrow \ket{5D_{3/2}, F'' = 4}$ transition is denoted $\Delta$, as indicated in Fig.~\ref{fig1}~(b). We choose $\delta$ sufficiently large such that the observed spectra are the result of pure two-photon transitions without significant influence of the $5P_{1/2}$ state's HF structure ($\sim 361.6$~MHz~\cite{SteckRb85}). Detailed schematics of cavity-mode stabilization and OPLLs are provided in~\cite{cardman2021}. 

After 10~ms of OL loading (full OL trap depth for atoms in the $5S_{1/2}$-level during the loading stage is $U_{latt}\approx h\times15$~MHz), the OL power is ramped down by means of an acousto-optical modulator [not shown in Fig.~\ref{fig1}~(a)] to a variable level, during which the $\sim$500~ns-long laser excitation is performed [see timing diagram in Fig.~\ref{fig1}~(c)]. The ions produced by PI due to 1064~nm light are guided through the electric-potential landscape applied via 6 electrodes [not shown in Fig.~\ref{fig1}~(a)] that surround the optical cavity to a multichannel plate (MCP) detector. Ion spectra at each power level of 1064~nm light are recorded for the subsequent analysis of the ac Stark shifts.

The ion signal observed at the lowest powers of 1064~nm light is, in part, attributed to Penning ionization, a process that has been found to be very effective in collisions of atoms in the $5D_{3/2}$ state~\cite{Barbier1, Barbier2}. Investigation of Penning and other ionization processes~\cite{Cheret_1982, barbier3} using setups similar to ours is beyond the scope of the present work but could be a topic of future studies.

\subsection{1064~nm power calibration}
\label{subsec:powercal}

\begin{figure}[b!]
 \centering
  \includegraphics[width=0.46\textwidth]{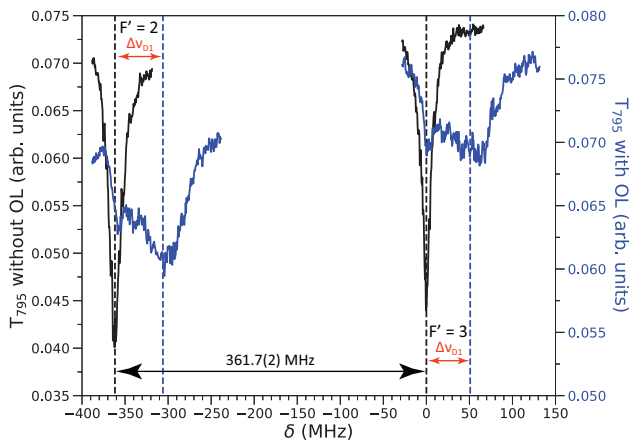}
  \caption{(Color online) Transmission of 795~nm beam, $T_{795}$, as a function of detuning, $\delta$, without and with OL used for the calibration of $I_{1064}$ using Eq.~\ref{eq:D1Shift}. Dashed vertical lines denote transmission minima occurring due to absorption of 795~nm light by atoms. The signal minima on the dashed lines at $\delta=0$ and -362~MHz are due to atoms that are not in the OL; these atoms have no ac shift. The broader signal minima, marked by the other pair of dashed lines, correspond to atoms trapped near the intensity maxima of the OL; these atoms experience ac shifts due to the OL field. From analysis of the data used for this plot, $\Delta \nu_{D1} = 56(4)$~MHz, $P_{T,1064} = 152(9)$~mV for $F' = 2$ and $\Delta \nu_{D1} = 51(6)$~MHz, $P_{T,1064} = 160(6)$~mV for $F' = 3$.
  }
  \label{fig2}
\end{figure}

As in similar measurements~\cite{Dinneen1992, duncan2001prameasurement}, it is important to properly calibrate the 1064~nm light intensity at the location of the atoms, $I_{1064}$. This is accomplished by measuring the transmitted 1064~nm power after the cavity output port with a photo-diode [PD in Fig.~\ref{fig1}~(a)] that is connected to a transimpedance converter with a fixed gain (output voltage divided by photo-current). The transmitted power is proportional to the recorded voltage, denoted $P_{T, 1064}$. Due to the linearity of the electronic circuit and the proportionality between $I_{1064}$ and the power transmitted through the cavity, it is $I_{1064} = \gamma P_{T, 1064}$. 

To obtain the calibration factor $\gamma$, we measure 795-nm transmission spectra of the Rb D1 line in the absence of 762~nm light for several OL intensities, for which the corresponding voltages $P_{T,1064}$ are recorded. To avoid saturation, we decrease the 795-nm power beam to $\sim$50~nW as measured in front of the cavity. An avalanche photodetector [APD in Fig.~\ref{fig1}~(a)] installed after the cavity output port records the transmission of 795-nm light, $T_{795}$ (see Fig.~\ref{fig2} for an example). Appropriate dichroic mirrors [DM1 and DM2 in Fig.~\ref{fig1}~(a)] eliminate the 762~nm and 1064~nm beams from the 795~nm transmission signal. 
Using known dynamic scalar polarizabilities of $5S_{1/2}$~\cite{bonin, marinescu1994, arora2012} and $5P_{1/2}$ states~\cite{Neuzner2015} in $^{85}$Rb at 1064~nm, the following expression for the ac Stark shift of the D1 line, $\Delta \nu_{D1}$, follows from Eq.~\ref{eq:ACShift}:

\begin{eqnarray}
\Delta \nu_{D1} & = & -(\alpha^{(0)}_{5P_{1/2}} - \alpha^{(0)}_{5S_{1/2}})\frac{ I_{1064}}{2 h c \epsilon_0} \nonumber \\ 
~ & = &
-(\alpha^{(0)}_{5P_{1/2}} - \alpha^{(0)}_{5S_{1/2}})\frac{\gamma P_{T, 1064}}{2 h c \epsilon_0} 
\quad.
\label{eq:D1Shift}
\end{eqnarray}

\noindent The AC shifts $\Delta \nu_{D1}$ are measured by comparing $D1$-line spectra with the OL turned off and on, as shown in Fig.~\ref{fig2}. 
The measurements of $\Delta \nu_{D1}$ and corresponding transmitted-power readings $P_{T,1064}$ then allow us to determine the calibration factor $\gamma$.
As the $5P_{1/2}$ level has two HF states, we determine $\Delta \nu_{D1}$ for each of them by multi-peak Gaussian fits and use both values to extract $\gamma$ as a weighted average. This allows us to reduce the uncertainty of $\gamma$. The result for $\gamma$ and its uncertainty are shown in Table~\ref{tab:table1}.

The uncertainty in the recorded $P_{T,1064}$ values and the uncertainty of $\gamma$ lead to final uncertainties of $I_{1064}$, using error propagation~\cite{taylor}. These are the main sources of uncertainty in the value of $\alpha_{5D_{3/2}}$ obtained in the next section.

\subsection{Extracting $\alpha_{5D_{3/2}}$}
\label{subsec:extr}

\begin{figure}[t!]
 \centering
  \includegraphics[width=0.48\textwidth]{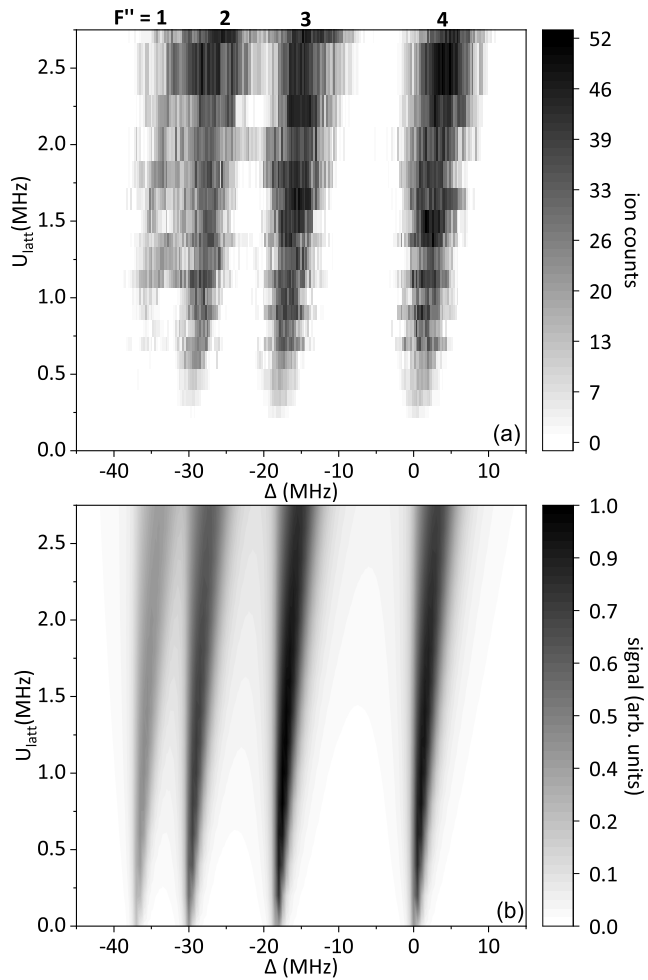}
  \caption{ Experimental ion spectra (a) and corresponding results of numerical simulations (b) as a function of two-photon detuning, $\Delta$, and OL depth, $U_{latt}$. Parameters used for (b) are $\alpha^{(0)}_{5D_{3/2}} = -499$~a.u., $\alpha^{(2)}_{5D_{3/2}} = 0$~a.u., and $\sigma = 44$~Mb.} 
  \label{fig3}
\end{figure}

We extract ac Stark shifts, $\Delta \nu_{F''}$, of different HF components of the $5D_{3/2}$ state by fitting Gaussian multi-peak profiles to the ion spectra. For the latter, we either acquire 5 to 10 scans and average them (at lowest 1064~nm power) or use single scans at good signal-to-noise ratios (at higher 1064~nm power, where PI is the most effective). Due to the HF structure of the Rb $5D_{3/2}$ state, the spectral lines from the two lowest HF states, $F'' = 1$ and $F'' = 2$ coalesce at lattice depths $U_{latt} \gtrsim h \times 1$~MHz (see Fig.~\ref{fig3}). Therefore, we restrict our data analysis to the $F'' = 3$ and $F'' = 4$ HF states, which are split by 18.6~MHz~\cite{nez1993optcomoptical} and shift linearly with $I_{1064}$:

\begin{equation}
\Delta \nu_{F''} = \beta_{F''} I_{1064},
\label{eq:5DShift}
\end{equation}

\noindent with a fitting parameter $\beta_{F''}$ that is related to $\alpha_{5D_{3/2}}$ as shown below in Eq.~\ref{eq:polformula}. Since the measurements are performed at low intensities of 1064~nm light and at large $\delta$, the excitation of the atoms in the OL field is an off-resonant two-photon excitation in which the intermediate $5P_{1/2}$ states do not become populated. Hence, ac-Stark-shifts of the $5P_{1/2}$ states caused by the OL do not enter into Eq.~\ref{eq:5DShift}, and the polarizability of the $5P_{1/2}$-state is not explicitly required. 
[Note that the experimental intensity calibration factor $\gamma = I_{1064}/P_{T,1064}$, which is critical to extract the $5D_{3/2}$ polarizability, does require the polarizability of the $5P_{1/2}$-state.] With the experimental calibration factor $\gamma = I_{1064}/P_{T,1064}$, and utilizing the finding that the tensor polarizability $\alpha^{(2)}_{5D_{3/2}}$ is too small to produce measurable effects~\cite{cardman2021}, Eqs.~\ref{eq:ACShift} and~\ref{eq:5DShift} yield

\begin{equation}
\alpha^{(0)}_{5D_{3/2}, F''} = \alpha^{(0)}_{5S_{1/2}} - 2hc\epsilon_0\beta_{F''}
\quad .
\label{eq:polformula}
\end{equation}

\noindent Since we analyze ac shifts of the two HF states $F''=3$ and $4$, the final measurement result $\alpha_{5D_{3/2}}$ is reported in Table~\ref{tab:table1} as a weighted average over two values.

\begin{table}[t!]
\caption{\label{tab:table1} Summary of quantities used in the data analysis and the final results.}
\begin{ruledtabular}
\begin{tabular}{l|c|c}
    Quantity & Value & Source\\
    \hline
    $\gamma$ & 37(3)~GW/m$^2$/V & This experiment \\ 
    $\alpha^{(0)}_{5S_{1/2}}$ & 687.3(5)~a.u. & \cite{arora2012} \\ 
    $\alpha^{(0)}_{5P_{1/2}}$ & -1226(18)~a.u. & \cite{Neuzner2015} \\
    $\beta_{F'' = 3}$ & 5.59(39)~MHz/GW/m$^2$ & This experiment\\
    $\beta_{F'' = 4}$ & 5.54(39)~MHz/GW/m$^2$ & This experiment\\
    $\alpha^{(0)}_{5D_{3/2}, F'' = 3}$ & -505(84)~a.u. & Eq.~\ref{eq:polformula}\\
    $\alpha^{(0)}_{5D_{3/2}, F'' = 4}$ & -494(83)~a.u. & Eq.~\ref{eq:polformula}\\
    $\alpha^{(0)}_{5D_{3/2}}$ & -499(59)~a.u. & Weighted average\\
\end{tabular}
\end{ruledtabular}
\end{table}

\section{Results and Discussion}
\label{sec:res}
\subsection{Determination of $\alpha_{5D_{3/2}}$}
\label{subsec:polresult}

\begin{figure}[t!]
 \centering
  \includegraphics[width=0.48\textwidth]{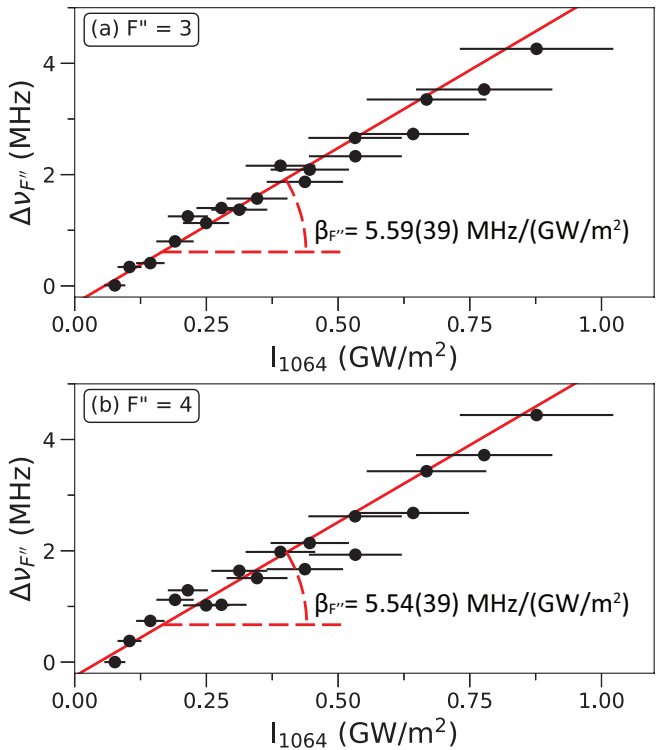}
  \caption{(Color online) ac Stark shifts, $\nu_{F''}$, vs $I_{1064}$ for $F'' = 3$ (a) and $F'' = 4$ (b). Dots with error bars are from experimental data. The solid lines are linear fits according to Eq.~\ref{eq:5DShift}.} 
  \label{fig4}
\end{figure}

The obtained ion spectra at calibrated $U_{latt} = \frac{\alpha_{5S_{1/2}}}{2 c \epsilon_0} \gamma P_{T,1064}$
are shown in Fig.~\ref{fig3}~(a). In a first cursory analysis, we compare the experimental spectra with a numerical simulation.
The close resemblance of the experimental data with the results of the numerical simulations in Fig.~\ref{fig3}~(b) is apparent. We attribute the larger broadening of the experimental spectra to the combined effects of the linewidths of the excitation lasers (both on the order of several 100~kHz), OL power fluctuations in the field enhancement cavity, and the MOT magnetic fields. The Penning ionization process mentioned in Sec.~\ref{subsec:extr}, which is not included in our simulation, may enhance the experimental level broadening at low lattice depths $U_{latt}$. 
Moreover, the dominance of light-shift effects in the present system over Doppler effects has led us to adopt a ``frozen'' Boltzmann distribution of atoms in the simulation, which does not account for atomic-motion effects in the OL. 
Apart from the fact that the $F'' = 1$ HF component is somewhat weaker in the experiment than in the simulation,  experimental and simulated data exhibit good qualitative agreement. 

For a quantitative analysis, we extract $\nu_{F''}$ for the $F''=3$ and 4 HF components and follow the procedure described in Sec.~\ref{subsec:extr}. The results (dots) together with the fitting to Eq.~\ref{eq:5DShift} (solid lines) are shown in Figs.~\ref{fig4}~(a) and~(b) for $F'' = 3$ and $F'' = 4$, respectively. The horizontal error bars are from the calibration of $I_{1064}$, as explained in Sec.~\ref{subsec:powercal}, while the vertical error bars (which are smaller than the dots) reflect statistical uncertainties from the fits to the experimental spectra. It is apparent from Figs.~\ref{fig4}~(a) and~(b) that the error bars due to the $I_{1064}$ calibration is the dominant source of uncertainty on $\beta_{F''}$ (see Table~\ref{tab:table1}). We use only the two upper HF levels $F''=3$ and $4$ for the data analysis because only these levels are  well-discerned throughout the range of the applied $I_{1064}$-values. The lower levels $F''=1$ and $2$ coalesce already at moderate $I_{1064}$ due to their small HF splitting (see Fig.~\ref{fig3}), making the spectral fits unstable for those levels.

The extracted values for $\alpha_{5D_{3/2}}$ of the $F'' = 3$ and 4 HF components as well as their weighted average are listed in Table~\ref{tab:table1}. The final result, $-499\pm 59$~a.u., agrees within the uncertainty range with the previously reported value~\cite{cardman2021}. The final uncertainty in the present paper is larger by a factor of $\approx$~3.5 due to the sensitivity of the result to the calibration error in $\gamma$ for the 1064~nm light intensity.

\subsection{Discussion of the result for $\alpha_{5D_{3/2}}$}
\label{subsec:polresultdisc}

The value of the present $\alpha_{5D_{3/2}}$-measurement relies on the fact that it is obtained in a different intensity regime than that used in~\cite{cardman2021} (low- vs high-intensity limit). The present measurement is performed in the low-intensity regime, in which laser intensities are on the same order of magnitude as in contemporary cold-atom research~\cite{AndersonSE2011, grossreview2017,  chomazarxiv2022, cardman2020}. 
The present measurement and the measurement in~\cite{cardman2021} exhibit different sets of systematic uncertainties. The observed agreement in $\alpha^{(0)}_{5D_{3/2}}$-values, within uncertainty limits, confirms the earlier result~\cite{cardman2021} and asserts 
that there are no critical omissions in the systematic effects affecting either of the measurements.

It is noteworthy that a previous theoretical analysis in~\cite{cardman2021} suggests that the dynamic \textit{tensor} polarizability of the $5D_{3/2}$ state, $\alpha^{(2)}_{5D_{3/2}}$ is very small.
According to Eq.~\ref{eq:poltotal}, a substantial $\alpha^{(2)}_{5D_{3/2}}$  would cause 
a dependence of the ac shifts in Figs.~\ref{fig3} and~\ref{fig4} and of the $\beta_{F''}$-values in Table~\ref{tab:table1}
on the upper HF state $F''$. Further, a substantial $\alpha^{(2)}_{5D_{3/2}}$ would cause line splittings of the $F''$ levels in Figs.~\ref{fig3} and~\ref{fig4}. The absence of such evidence in the measurements is consistent with $\alpha^{(2)}_{5D_{3/2}} \sim 0$. An actual measurement of $\alpha^{(2)}_{5D_{3/2}}$ is hampered, if not made impossible, by the line broadening in Figs.~\ref{fig3} and~\ref{fig4}, which is almost entirely due to PI of atoms by the 1064-nm OL light.

\begin{figure}[b!]
 \centering
  \includegraphics[width=0.48\textwidth]{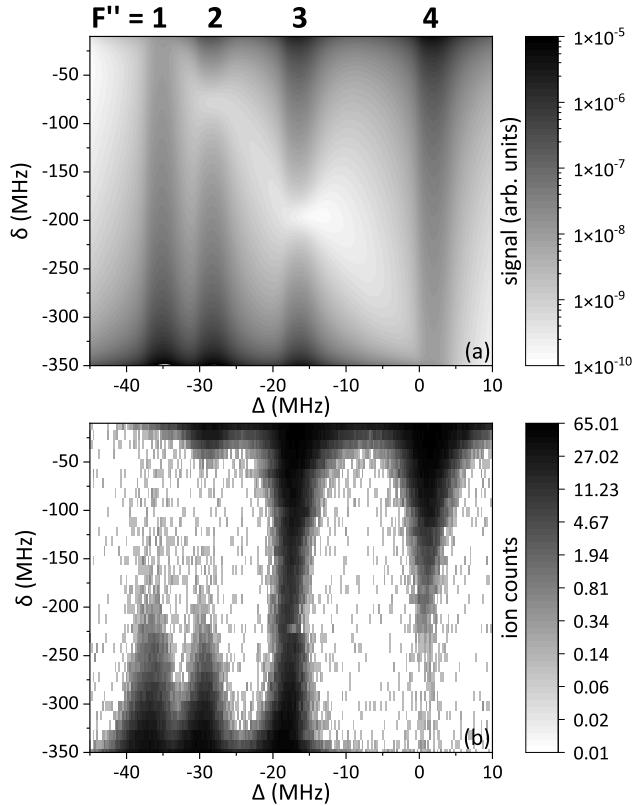}
  \caption{ Simulated (a) and experimental (b) $5D_{3/2}$ ion spectra as a function of lower, $\delta$, and upper, $\Delta$, transition detunings at $U_{latt}\approx h \times 2$~MHz. See text for discussion.} 
  \label{fig5}
\end{figure}

\subsection{Quantum interference at $\delta < 0$}
\label{subsec:qi}

The measurements reported above are performed at a large positive detuning, $\delta$, from the $\ket{5S_{1/2}, F = 3} \rightarrow \ket{5P_{1/2}, F' = 3}$ transition to avoid any influence of the Rb $5P_{1/2}$-state HF structure on the result. In order to explore how the $5P_{1/2}$ HF structure 
affects the two-color $5D_{3/2}$ excitation spectra, we have performed additional simulations at $U_{latt} = h \times 2$~MHz with $\delta$ varied between -10~MHz and -350~MHz, scanning most of the range between $F'=2$ and $3$. The obtained simulated map is shown in Fig.~\ref{fig5}~(a) as a function of $\delta$ and $\Delta $ as defined in Fig.~\ref{fig1}~(b). 
In this representation, the $F'$-resonances would appear as signal maxima lined up along horizontal lines at $\delta = -362~$MHz and 0 [just off-scale in Fig.~\ref{fig5}~(a)], while the $F''$-resonances appear as vertical bars at $\Delta$-values given by the respective $F''$ HF shifts.
The signal in Fig.~\ref{fig5}~(a) is characterized by some amount of line broadening along $\Delta$ due to lattice-induced PI, which sets a floor of several MHz for the linewidth in $\Delta$. When approaching $\delta=-362$~MHz and $0$ along the $\delta$-axis, the signals become stronger and substantially broadened due to the near-resonant excitation of $5D_{3/2}$-atoms through the intermediate $F'=2$ and $3$ levels, respectively. 
In Fig.~\ref{fig5}~(a) we further observe signal dropouts on the $F''=2$ and $F''=3$-lines 
near $\delta \approx -75$~MHz and $\approx -200$~MHz. The dropouts are due to quantum interference in the excitation amplitudes of the $F'' = 2$ and $F'' = 3$ HF states. As $\delta$ is varied, the two-photon Rabi frequencies of the excitation channels $\ket{5S_{1/2}, F = 3} \rightarrow \ket{5P_{1/2}, F' = 2} \rightarrow \ket{5D_{3/2}, F''}$ and $\ket{5S_{1/2}, F = 3} \rightarrow \ket{5P_{1/2}, F' = 3} \rightarrow \ket{5D_{3/2}, F''}$ vary in relative strength due to their different, $\delta$-dependent intermediate detunings. For $F''=2$ and 3 both two-photon Rabi frequencies are non-zero and have opposite signs in the displayed $\delta$-range. Hence, at certain $\delta-$values destructive interference must occur, causing the signal dropouts in  Fig.~\ref{fig5}~(a).

In corresponding experimental data shown in Fig.~\ref{fig5}~(b) we generally observe the simulated behavior. Notable differences include the large strength of the $F''=3$-signal throughout the experimental data, as well as differences in the prominence of the signal dropouts caused by the destructive quantum interference. The discrepancies may be caused by optical-pumping and atomic-motion effects, which were both not included in the simulation. An improved numerical model, which is beyond the scope of the present work, may include the center-of-mass dynamics of the atoms on the lattice potentials.
Certain techniques from computational molecular dynamics, such as fewest-switches surface-hopping method~\cite{Tully1990, craig2005}, could be employed for this purpose in future studies.

\section{Conclusion}
\label{sec:concl}
In summary, we have reported a measurement of the dynamic polarizability of the Rb $5D_{3/2}$ state in 1064~nm optical fields in the ``Zeeman'' regime, where the ac shifts are less than the $5D_{3/2}$ hyperfine splittings.
Our results show that the ac shifts of the $5D_{3/2}$ HF states approach the HF splittings already at moderate 1064-nm intensities, equivalent to commonly used optical-trap depths. 
Owing to the large $5D_{3/2}$ PI cross-section at 1064-nm wavelength, the spectroscopic lines exhibit significant broadening effects. 
In metrological applications~\cite{hilico1998epjapmetrological, terra2016apbultra, martin2019, quinn2003metpractical} of cold $5D_{3/2}$ Rb atoms in 1064-nm laser traps, both ac shifts and PI broadening will have to be accounted for and minimized as needed. Future studies could be directed towards precise determination of the tensor contribution to the total polarizability, which was neglected in our analysis. To characterize the response of the $5D_{3/2}$ state to optical fields at a precision sufficient for this purpose, such studies could be performed at wavelengths longer than the PI threshold (1251.52~nm for Rb $5D_{3/2}$). Ac shifts and PI broadening should also be taken into account when using the Rb $5D_{3/2}$ state as an intermediate excitation level in experiments on Rydberg atoms and molecules~\cite{Moore2019a, duspayev2021, Younge2010, cardman2021njp}. Conversely, PI of Rb $5D_{3/2}$ atoms provides an efficient method to prepare cold ion clouds for research on non-neutral plasmas, allowing studies of highly-excited Rydberg atoms immersed in such plasmas~\cite{seckerpra, ewald, weber2012, anderson2017, Ma20}.

\maketitle
\section*{ACKNOWLEDGMENTS}
This work was supported by the NSF Grant No. PHY-2110049. We would like to thank Dr. Yun-Jhih Chen, Dr. Xiaoxuan Han and Dr. Jamie MacLennan for useful discussions and initial experimental work.

\bibliography{references}
\end{document}